\documentclass[hyper]{JHEP3}

\input{epsf}
\usepackage{epsfig}
\usepackage{amssymb}
\usepackage{amsfonts}
\usepackage{amsbsy}
\usepackage{amsmath}

\usepackage{epstopdf}

\def\IC{\mathbb{C}}

\def\IZ{{\mathbb{Z}}}
\def\IR{{\mathbb{R}}}

\def\CN {{\cal N}}

\def\CF {{\cal F}}

\def\CP {{\cal P }}

\def\CO {{\cal O}}

\def\CH {{\cal H}}

\def\CQ{{\cal Q}}
\def\CU{{\cal U}}

\def\one{{\hbox{ 1\kern-.8mm l}}}

\newcommand\fro{{\overline{\underline{\Omega}}}}

\def\F{G}  

\title{Wall-crossing from supersymmetric galaxies}

\author{Evgeny~Andriyash$^1$, Frederik~Denef$^{2,3}$, Daniel~L.~Jafferis$^4$ and Gregory~W.~Moore$^1$\\
\\
${}^1$ NHETC and Department of Physics and Astronomy,
Rutgers University,\\
$^2$ Center for the Fundamental Laws of Nature,
Harvard University,\\
$^3$ Institute for Theoretical Physics,
University of Leuven, \\
$^4$ School of Natural Sciences, Institute for Advanced Study, Princeton.\\
\\
{\tt andriyas@physics.rutgers.edu, denef@physics.harvard.edu,\\
jafferis@ias.edu, gmoore@physics.rutgers.edu} }

\abstract{We give an elementary physical derivation of the Kontsevich-Soibelman wall crossing formula, valid for any theory with a 4d $\CN=2$ supergravity description. Our argument leads to
 a slight generalization of the formula, which relates monodromy to the BPS spectrum.
}

\preprint{RUNHETC-2010-15}

\begin{document}


\section{Introduction}

In \cite{KS-1} Kontsevich and Soibelman proposed a remarkable wall crossing formula for BPS indices.
In this note we show that this formula can be derived in an elementary way from the halo picture of BPS bound states in supergravity \cite{Denef:2002ru,Denef:2007vg}.

The basic strategy we follow is similar to that of \cite{Gaiotto:2010be}, which gave a proof of the (motivic)
KS wall-crossing formula in the context of $\CN=2$ \emph{field theory}. The essential physical
idea used halo configurations of  particles bound to line operators.  Our analysis will generalize this idea to gravity, without introducing external objects such as line operators. The surrogate for the line operator
 will be an infinitely massive BPS black hole, to which the BPS objects of interest are bound. The physical cartoon to have in mind is that of a galaxy with a supermassive black hole at its center, where the BPS objects of interest are the solar systems orbiting around it. These galactic configurations exhibit jumping phenomena when dialing the moduli: when crossing certain walls, halos of objects of a particular charge can be pushed out to infinity or conversely come in from infinity. The generating function for the BPS indices of these galactic bound states transforms in a simple way when such a wall is crossed, by the action of a certain operator on the generating function, which follows directly from the simple halo wall crossing formula (a.k.a.\ the semiprimitive wall crossing
 formula) of \cite{Denef:2007vg}. Collections of walls intersect on real codimension two loci, together also with marginal stability walls for the individual solar systems. Circling around these intersection loci will produce a sequence of wall crossing operations on the generating function. For a contractible loop in moduli space, the product of these operators must be the identity. This turns out to be nothing but the KS formula. For a noncontractible
 loop in moduli space we find a   generalization of the KS formula.

 We refer to the companion paper \cite{BST} for further background and examples. Our notation
 follows \cite{BST} which in turn follows the notation of \cite{Denef:2007vg}.

\section{BPS galaxies and the halo wall crossing operator}\label{sec:WCoperators}

A halo is a BPS configuration consisting of an arbitrary number $N$ of particles with electromagnetic
 charge proportional to a primitive charge $\gamma$ surrounding a core of charge $\Gamma$.
 For simplicity of exposition (only!) we will initially consider only halo particles of charge $\gamma$.
 The charges are valued in a symplectic lattice $L$.
The equilibrium distance $R$ between core and halo particles is given by \cite{Denef:2000nb}
\begin{equation} \label{Req}
 R = \frac{\langle \gamma,\Gamma \rangle}{2 \, {\rm Im}( e^{-i \alpha} Z_\gamma)} \, ,
\end{equation}
where $\langle \gamma, \Gamma \rangle$ is the electric-magnetic symplectic product of $\gamma$ and $\Gamma$, $Z_\gamma$ is the central charge of $\gamma$, measured at spatial infinity (where the vector multiplet moduli are set at $t=t_\infty$), and $\alpha = \arg Z_{\Gamma+N \gamma}$. A necessary condition for existence is $R>0$. When the phases of the central charges of the core and halo line up, i.e.\ $\arg Z_\Gamma = \arg Z_\gamma = \alpha$, the radius diverges and the halo decays. Both core and halo particles can in turn be composites. The above formula for the equilibrium distance still holds as long as $R$ is much larger than the size of these composites.

In the limit $R \to \infty$, the halo particles can be considered to be noninteracting electric point particles, confined to a sphere threaded by a uniform magnetic flux. The supersymmetric one particle ground states are given by the lowest Landau levels, and the $N$-particle halo states are constructed from those as an $N$ particle Fock space $\CF_\Gamma(N \gamma)$ \cite{Denef:2002ru,Denef:2007vg}. We denote the Witten index of these halo states by\footnote{The indices depend on the background moduli $t_\infty$. For notational compactness we will sometimes
suppress this dependence.  }
\begin{equation}
 \Omega_{\Gamma}^{\rm Fock}(N \gamma;t_\infty) \equiv {\rm Tr}_{\CF_\Gamma(N \gamma)} (-1)^F \, .
\end{equation}
For $N=1$, we have $\Omega^{\rm Fock}_\Gamma(\gamma) = |\langle \gamma,\Gamma \rangle| \Omega(\gamma)$. Here  $\Omega(\gamma)$ is the usual $\CN=2$ BPS index, and $|\langle\gamma,\Gamma\rangle|$ is the lowest Landau level degeneracy factor. For general $N$ it is convenient to define a generating function. Introduce formal variables $X_i$, $ i= 1, \dots, {\rm rank}L$, and write $X^\Delta := \prod_i X_i^{\Delta_i}$ for a charge $\Delta$ with components $\Delta_i$ with respect to some chosen basis for $L$.  Then the generating function is
\begin{equation} \label{FockZ}
 \F_\Gamma^{\rm Fock}(X) := \sum_N \Omega^{\rm Fock}_\Gamma(N \gamma) \, X^{\Gamma + N \gamma} =
 \left(1 - (-1)^{\langle \gamma,\Gamma \rangle} X^\gamma \right)^{\Omega(\gamma) |\langle \gamma,\Gamma \rangle|} \,
  X^\Gamma
\end{equation}
This follows from standard Fock space combinatorics \cite{Denef:2007vg}.

In general $\Omega(\Gamma+N \gamma) \neq \Omega(\Gamma) \, \Omega_\Gamma^{\rm Fock}(N \gamma)$ in the full theory.
The reason is that   the true index $\Omega(\Gamma+N \gamma)$ in general gets contributions from many other configurations of charges summing up to the same total charge. For instance a core black hole of charge $\Gamma$ with two halo particles of charge $\gamma$ and a core black hole of charge $\Gamma+\gamma$ and one halo particle of charge $\gamma$ will both contribute to $\Omega(\Gamma+2 \gamma)$. At finite $R$, the corresponding Fock spaces can be expected to get mixed due to quantum tunneling between these configurations. Only the sum over all possible configurations is guaranteed to give a well defined index. Phrased differently, whereas the supersymmetric quantum mechanics of halo particles trapped in their potential minimum at finite $R$ is a well-defined closed system in perturbation theory, nonperturbative tunneling between this minimum and the minimum corresponding to merging with the black hole core causes the wave function of the halo configuration to ``leak out'' and mix with configurations with different core black hole charges. It is no longer a well-defined closed system.

The leaking can be prevented, however, by taking the limit of infinite core black hole size, as black hole tunneling is generically exponentially suppressed in the size of the black hole. This is entirely an entropic effect. For example the amplitude for fragmentation of an extremal Reissner-Nordstr\"om black hole of charge $Q=Q_1+Q_2$ into black holes of charge $Q_1$ and $Q_2$ --- a process unobstructed by any potential barrier --- is nevertheless suppressed as $e^{-\frac{1}{2} \Delta S}$ where $\Delta S = \pi Q^2 - \pi Q_1^2 - \pi Q_2^2 = 2 \pi Q_1 Q_2$ \cite{Maldacena:1998uz}. Therefore in the $Q \to \infty$ limit, taking into account charge quantization, the extremal RN black hole becomes absolutely stable; there is no more mixing with fragmented configurations. Stability of large black holes is a universal phenomenon --- even Schwarzschild black holes stop radiating and become stable in the infinite size limit.

Thus, we will consider configurations of BPS objects orbiting around a supermassive black hole core of charge $\Gamma_c$, where we eventually send $\Gamma_c \to \infty$ while keeping the total charge of the objects in the orbits finite. The objects themselves can be multicentered BPS bound states. We can loosely think of this system as a galaxy consisting of many solar systems orbiting around a supermassive black hole, and we therefore refer to these objects as ``BPS galaxies''. The simplest situation is when we have a single halo of particles of charge $\gamma$ around the hole, but we also allow multiple halos, or more general, non-halo configurations involving interacting solar systems with mutually nonlocal charges. So the most general BPS galaxy will be a complex multi-particle bound state, with potentially strong position-constraining interactions between neighboring solar systems, and intricate exchanges of suns and planets between different solar systems possible when dialing the moduli.

 To make this more precise, we have to specify more carefully how we take the limit $\Gamma_c \to \infty$. For our purpose of deriving the KS formula, it turns out to be convenient to single out a particular $U(1)$, give the core large electric and magnetic charges with respect to this $U(1)$, and keep the orbiting solar systems uncharged under this $U(1)$. More precisely, we choose a set of charges $C \equiv \{\Gamma_0,\Gamma_0',\gamma_c\}$ such that $\Gamma_0$ supports a single centered BPS black hole, $\langle \Gamma_0,\Gamma_0' \rangle \neq 0$, and $\langle \gamma_c,\Gamma_0 \rangle = 0 = \langle \gamma_c,\Gamma_0' \rangle$. We then set
\begin{equation} \label{GamLam}
 \Gamma_c = \Lambda^2 \Gamma_0 + \Lambda \Gamma_0' + \gamma_c \,
\end{equation}
and take $ \Lambda \to \infty$. The anisotropic scaling is chosen for reasons that will become clear later (see footnote \ref{whyscaling}). To avoid infinite lowest Landau level degeneracies, we restrict the charges $\gamma$ of the solar systems orbiting around this core to be orthogonal to both $\Gamma_0$ and $\Gamma_0'$, which means they are uncharged under the $U(1)$ associated to $\Gamma_0$ and $\Gamma_0'$. More formally, the sublattice of orbiting charges $L_{\rm orb}$ is thus
\begin{equation}
 L_{\rm orb}:= \{ \gamma \in L \, | \, \langle \gamma,\Gamma_0 \rangle  = 0 = \langle \gamma,\Gamma_0' \rangle \}.
\end{equation}
With this definition, we also have $\gamma_c \in L_{\rm orb}$.

The Hilbert space of BPS galaxies with core charge $\Gamma_c$ and total orbital charge $\Gamma_{\rm orb}$ has an overall factor corresponding to the internal states of the
 core black hole, which we can factor out to produce a factor space  $\CH_{\Gamma_c}(\Gamma_{\rm orb};t_\infty)$, which can be thought of as the Hilbert space of the orbiting solar systems in a background sourced by the core black hole. We obtain a closed supersymmetric quantum system with this Hilbert space provided there is no mixing between galaxies of different core charges, nor mixing with galaxies which do contain charges in orbit which are \emph{not} in the restricted lattice $L_{\rm orb}$. This turns out to be generically the case in the limit $\Lambda \to \infty$, essentially because such tunneling events are either infinitely entropically suppressed along the lines mentioned above, or infinitely suppressed because they require tunneling over infinite distances. We give detailed arguments for this in appendix \ref{sec:MultipleCore}, and prove that there is just one exception, namely when it so happens that the attractor point of $\Gamma_0$ lies on a locus with massless particles with charge in $L_{\rm orb}$, in which case there may be mixing between galaxies with cores differing by the charges becoming massless. This situation is nongeneric, and for the remainder of the paper we will assume this is not the case.

Thus, at fixed $\Gamma_{\rm orb}$, in the limit $\Lambda \to \infty$, we can define a proper Witten index for this supersymmetric closed system, which we call the ``framed'' BPS galaxy index, in analogy with the framed BPS indices of \cite{Gaiotto:2010be}:
\begin{equation}\label{eq:Framed-Def}
 \fro_{C}(\Gamma_{\rm orb};t_\infty) := \lim_{\Lambda \to \infty} {\rm Tr}_{\CH_{\Gamma_c}(\Gamma_{\rm orb};t_\infty)} \, (-1)^F  \, .
\end{equation}
Here $C \equiv \{\Gamma_0,\Gamma_0',\gamma_c \}$ is the set of charges determining the one parameter family $\Gamma_c(\Lambda)$ of core charges as in (\ref{GamLam}).
It will be useful to introduce the generating function of framed BPS indices:
\begin{equation}\label{eq:gen-fun}
\F_{C}(X;t_\infty):= \sum_{\Gamma_{\rm orb}\in L_{\rm orb}} \fro_{C}(\Gamma_{\rm orb};t_\infty) \, X^{\gamma_c + \Gamma_{\rm orb}}.
\end{equation}

The presence of singularities and associated monodromies gives rise to some subtleties, which we discuss in section \ref{sec:noncontr}. For the time being we simply assume we stay in a sufficiently small open set of moduli space, away from singular loci, in which case we can ignore these subtleties altogether.

The key observation that makes this construction useful is that although the generic BPS galaxy has a very complicated structure, its wall crossing behavior when varying $t_\infty$ is very simple. It is entirely governed by pure halo decays, since the galactic core black hole cannot decay and serves as a fixed, primitively charged center. Whenever the central charge $Z(\gamma)$ of some charge $\gamma$ supporting BPS states lines up with the total central charge $Z=Z(\Gamma_c) + Z(\Gamma_{\rm orb})$ of the galaxy, a halo of   objects with charge    $\gamma$ can be added or subtracted at spatial infinity. We again restrict to $\gamma \in L_{\rm orb}$. In the $\Lambda \to \infty$ limit the wall in moduli space where this happens is independent of the solar system charge, since in this limit $Z/Z(\Gamma_c) = 1$, so $\arg Z = \arg Z(\Gamma_c) = \arg Z(\Gamma_0)$ and we can set $\alpha = \alpha_0:= \arg Z(\Gamma_0;t_\infty)$ in (\ref{Req}).  Hence the wall of marginal stability for the halo is
\footnote{These are analogs of the ``BPS walls'' of \cite{Gaiotto:2010be} with $e^{i \alpha_0}$ playing the role of $\zeta$. However, an important difference is that now $e^{i \alpha_0}$ depends on $t_\infty$ and is only an independent variable to the
extent that $\Gamma_0$ is.}
\begin{equation} \label{Wg}
 W_\gamma = \{ t_\infty | \arg [e^{-i \alpha_0} Z(\gamma,t_\infty)] = 0  \} \, , \qquad \mbox{stable side:}  \quad \langle \gamma,\gamma_c + \Gamma_{\rm orb} \rangle \, {\rm Im}[e^{-i \alpha_0} Z(\gamma,t_\infty) ] > 0 \, .
\end{equation}
We will call these ``BPS walls.''

The part of the Hilbert space of all BPS galaxies with fixed core charge $\Gamma_c$ that jumps across a
 BPS wall $W_\gamma$ is given by the halo Fock space   described earlier, with an effective core charge $\Gamma$, as seen by this halo, given by the \emph{total} interior galactic charge $\Gamma=\Gamma_c+\Gamma_{\rm orb}$ enclosed by the halo. The corresponding transformation of the framed galactic indices can therefore be inferred from (\ref{FockZ}). Roughly speaking, the terms in the generating function $\F_{C} $ in \eqref{eq:gen-fun} get multiplied by the factor appearing in (\ref{FockZ}). However, as we have just explained,  the effective $\Gamma$ appearing in (\ref{FockZ}) depends on $\Gamma_{\rm orb}$
 and hence is different for the different terms in $\F_{C}$, and so the multiplication factor will be different. This is easily formalized by introducing a linear operator $D_\gamma$ acting on monomials $X^\delta$ by pulling down the symplectic product:
\begin{equation}
 D_\gamma X^\delta := \langle \gamma,\delta \rangle X^\delta \, .
\end{equation}
With this and an eye on (\ref{FockZ}), we define the following operator acting on polynomials in $X$:
\footnote{We remark that the operators $\tau_{\gamma}:=(-1)^{D_\gamma} X^\gamma$ satisfy
$\tau_{\gamma} \tau_{\gamma'} = (-1)^{\langle \gamma, \gamma'\rangle} \tau_{\gamma+\gamma'}$,
and hence the operators $\tau_\gamma$ provide a natural quadratic refinement of the
mod-two intersection form, a point which aficionados of the KSWCF will surely appreciate.
(A related point was made in equations (3.27)-(3.29) of \cite{Gaiotto:2010be}.)}
\begin{equation}
 T_\gamma := \left(1 - (-1)^{D_\gamma} X^\gamma \right)^{D_\gamma} \, .
\end{equation}
Notice that this operator effectively acts as a diffeomorphism on the coordinates $X^i$. The transformation of the generating function when crossing the wall $W_\gamma$ in the direction of \emph{increasing} $\arg [Z_\gamma e^{-i \alpha_0} ]$ is then
\begin{equation} \label{hwcf}
 \F_{C}(X) \to U_\gamma(t) \,  \F_{C}(X) \, , \qquad
 U_\gamma(t) := \prod_{k \in \IZ^+} T_{k \gamma}^{\Omega(k \gamma;t)} \, ,
\end{equation}
where we made the dependence on the point $t$ where the wall is crossed explicit. We take $\gamma$ to be primitive. The product over $k$ comes from the fact that the walls $W_{k \gamma}$ coincide. (Thus, we have now
 relaxed our initial assumption that only halo particles of primitive charge $\gamma$ enter.) To check that this formula is correct when going in the direction of increasing $\arg [Z_\gamma e^{-i\alpha_0}]$, note that on the part of the generating function for which $D_\gamma > 0$, going in this direction means by (\ref{Wg}) going from the unstable to the stable side, and vice versa for the $D_\gamma<0$ part. Therefore, the wall crossing formula should multiply the $D_\gamma>0$ terms by halo factors (\ref{FockZ}), and conversely remove such factors from the $D_\gamma<0$ part (or alternatively add such factors when the inverse operation is performed, corresponding to decreasing $\arg [Z_\gamma e^{-i\alpha_0}]$). This is indeed implemented by the fact that we dropped in (\ref{hwcf}) the absolute value signs appearing in the exponent of (\ref{FockZ}).

 Finally, we come to the central formula of this Letter. Consider a
closed contractible loop $\CP$ in moduli space (noncontractible loops will be discussed in section \ref{sec:noncontr}). Along this loop, the generating function $\F_{C}$ will undergo a sequence of wall crossing operations $U_{\gamma_i}(t_i)$.
Since $\CP$ is contractible, the composition of these operations must act trivially on $\F_{C}$, for any choice of $\gamma_c$ and starting point $t$:
\begin{equation} \label{eq:trivmonF1}
  \prod_i U_{\gamma_i}(t_i) \cdot \F_{C}   = \F_{C}  \, ,
\end{equation}
where the product is ordered according to the sequence of walls crossed: points crossed
later in the path are placed to the left. At the core attractor point $t_*(\Gamma_c)$ there are no multicentered bound states involving $\Gamma_c$, and hence no BPS galaxies. So at this point we have simply
\begin{equation} \label{attrval}
 \F_{C}(X)|_{t_\star(\Gamma_c)} = X^{\gamma_c} \, .
\end{equation}
Starting from this expression, the wall crossing formula (\ref{hwcf}) uniquely determines all framed galactic indices given all $\Omega(k\gamma)$. This shows that $\F_{C}$ is well defined as a function to the extent that the wall crossing factors are. (It is conceivable that a dense set of BPS walls can lead to an ill-defined expression.) Furthermore by varying $\gamma_c$ we can generate as many independent functions $\F_{C}(X)$ as there are independent variables $X_k$ associated to charges in $L_{\rm orb}$.
 \footnote{This corresponds to the condition, discussed in \cite{Gaiotto:2010be},
 that there are ``enough'' line operators to deduce the KSWCF. } This in combination with the fact that the wall crossing operators $U_\gamma$ act as diffeomorphisms implies that the product of the sequence of halo wall crossing operators around a contractible loop must be the identity
\begin{equation} \label{eq:trivmon}
  \prod_i U_{\gamma_i}(t_i) = 1 \, .
\end{equation}
We will prove in detail in the next section that this is in essence equivalent to the KS wall crossing formula.

\section{Derivation of the KS formula}\label{sec:KSformula}

\begin{figure}[htp]
\centering
\includegraphics[scale=0.3,angle=0,trim=0 0 0 0]{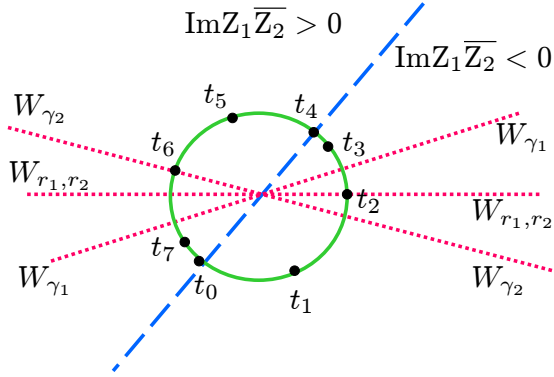}
\caption{This shows the neighborhood $\CU$ in the   normal bundle to $W_{\gamma_1}\cap W_{\gamma_2}$.
The wall of marginal stability is given by ${\rm Im}[Z(\gamma_1;t)\overline{Z(\gamma_2;t)}] =0$
since ${\rm Re}[Z(\gamma_1;t)\overline{Z(\gamma_2;t)}] $ is nonzero throughout $\CU$.
We choose the ordering of  $\gamma_1, \gamma_2$ so that $W_{\gamma_1}$ is counterclockwise from $W_{\gamma_2}$
with opening angle smaller than $\pi$.
Then the BPS walls $W_{r_1\gamma_1 + r_2 \gamma_2}$ are ordered so that increasing  $r_1/r_2$ gives walls
in the counterclockwise direction.  We
consider a path $\CP$ in $\CU$ circling the origin in the counterclockwise direction. The central charges of vectors $r_1\gamma_1 + r_2 \gamma_2$ with $r_1, r_2 \geq 0$ at representative points $t_0, \dots, t_7$ along $\CP$
are illustrated   in the next figure.   }
\label{fig:Walls}
\end{figure}

\begin{figure}[htp]
\centering
\includegraphics[scale=0.3,angle=0,trim=0 0 0 0]{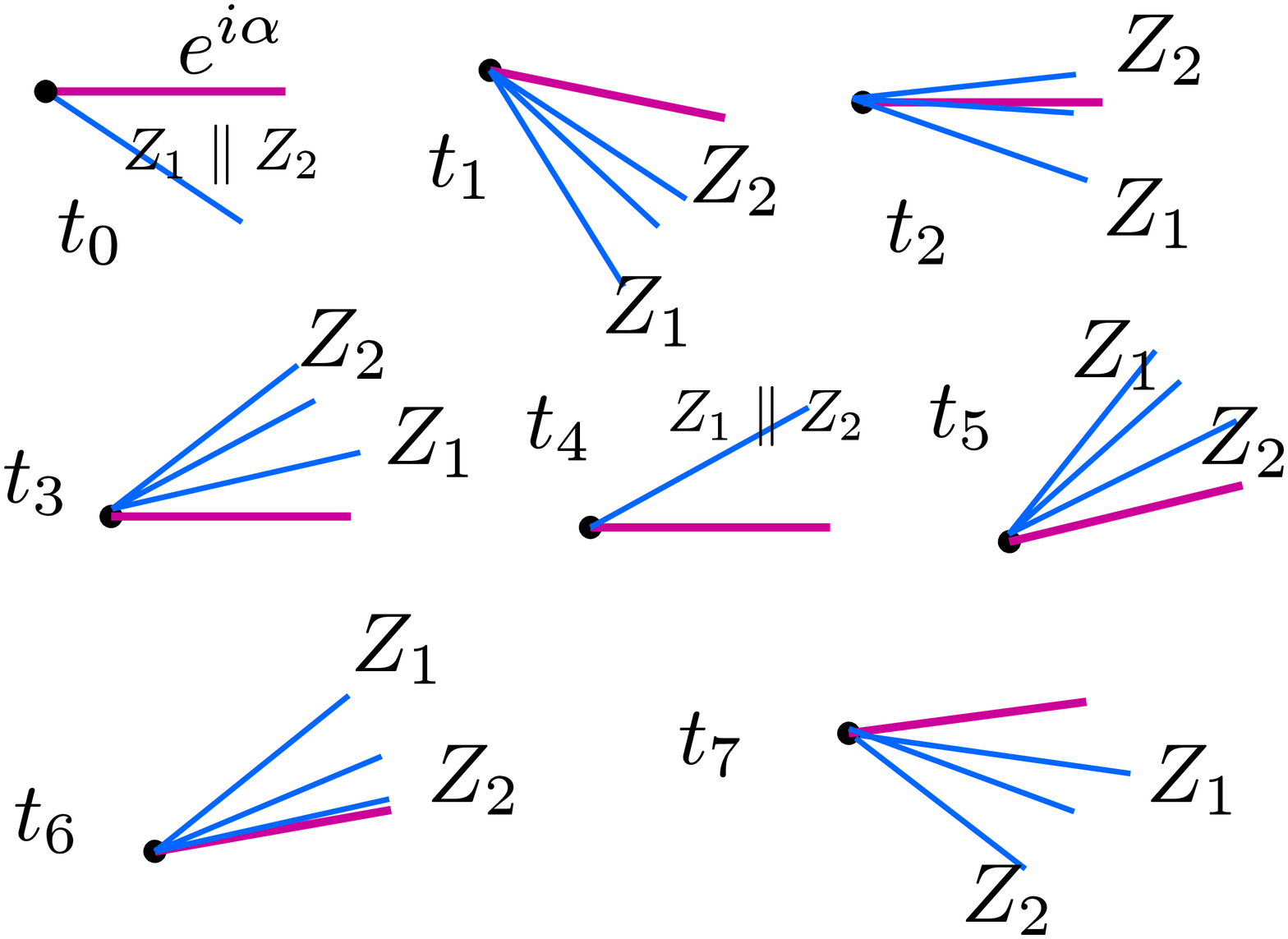}
\caption{As $t$ moves along the path $\CP$ the central charges evolve as in this figure. Note that
${\rm Im}(Z_1 \overline{Z_2}) >0$ means that $Z_1$ is counterclockwise to $Z_2$ and rotated by a phase
less than $\pi$. In that case the rays parallel to $r_1 Z_2 + r_2 Z_2$ for $r_1, r_2\geq 0$ are
contained in the cone bounded by $Z_1 \IR_+$ and $Z_2\IR_+$,
 and ordered so that increasing $r_1/r_2$ corresponds to moving counterclockwise.
When $t$ crosses the marginal stability wall the cone collapses and the rays reverse order. As $t$ moves in
the region $t_2$ the quantity ${\rm arg} [Z_\gamma e^{-i \alpha_0}]  > 0$ is increasing for all
$\gamma_{r_1,r_2}$  with  $r_1, r_2\geq 0$ while at the point $t_6$ the argument is decreasing.  }
\label{fig:CentralCharges}
\end{figure}

We now demonstrate that when $\CP$ is a small contractible loop intersecting a wall of marginal stability
the Kontsevich-Soibelman wall crossing formula is a consequence of  \eqref{eq:trivmon}.
Let us therefore consider two  mutually nonlocal charges $\gamma_1, \gamma_2$ and a generic {\it non-singular} point $t_{ms}\in MS(\gamma_1,\gamma_2)$ where both central charges are nonzero and $\gamma_1,\gamma_2$ support BPS states.
Using the attractor equation it is easy to
show that we can always find a $\Gamma_0$ (and hence a phase $\alpha_0$) so that $\Gamma_0$ supports single-centered
black holes and $t_{ms}$ lies on the intersection of BPS walls $W_{\gamma_1} \cap W_{\gamma_2}$.
\footnote{We can take for example $\Gamma_0 \equiv -2 {\rm Im}[\bar{X} \Omega^{(3,0)}]$, where $\Omega^{(3,0)}$ is the holomorphic 3-form evaluated at $t_{ms}$ and $X$ is an arbitrary complex constant with $\arg X \equiv \arg Z_1 = \arg Z_2$. This $\Gamma_0$ has a regular attractor point, namely $t_{ms}$, because the equation we used to define $\Gamma_0$ is nothing but the attractor point equation. Taking the symplectic product of this equation with $\gamma_1,\gamma_2$ shows that $\langle \gamma_1,\Gamma_0 \rangle = 0 = \langle \gamma_2,\Gamma_0 \rangle$. Taking the symplectic product with $\Omega^{(3,0)}$ shows that $X=Z(\Gamma_0:t_{ms})$, so, as we wished, the central charges line up at $t_{ms}$. Although $\Gamma_0$ will in general not be quantized, this is acceptable since all we care about in the end is the limit $\Lambda \to \infty$.
}
This intersection
is real codimension two in moduli space and we now consider a small neighborhood $\CU$ of $t_{ms}$ so that the only
other BPS walls $W_{\gamma'}$ passing through $t_{ms}$ arise from charges of the form
$\gamma' = r_1 \gamma_1 + r_2 \gamma_2$ for rational
$r_1, r_2$.
We will denote charges of this form by $\gamma_{r_1,r_2}$. Since the point $t_{ms}$ is non-singular a loop around it is contractible and (\ref{eq:trivmon}) holds.

Below we will argue that,  perhaps after choosing suitable linear combinations, we can assume that
the only populated charges of type $ \gamma_{r_1,r_2}$ in $\CU$  in fact have
$(r_1, r_2) \in \IZ^2$ with $r_1,r_2$ both $\geq 0$ or both $\leq 0 $. We can order $\gamma_1, \gamma_2$
so that the configuration of BPS walls and the marginal stability wall are arranged as shown in Figure \ref{fig:Walls}.
Suppose we begin at the point $t_1$ and move along the path $\CP$ in the counterclockwise direction.
We first cross the BPS walls in the region ${\rm Im} Z_1\overline{Z_2} <0$
in order of increasing $r_1/r_2$ and \emph{increasing} ${\rm arg}[ Z_{\gamma_{r_1,r_2}} e^{-i \alpha_0}]$.
Then we cross in the region ${\rm Im} Z_1\overline{Z_2} >0$ again with increasing $r_1/r_2$ but now
this corresponds to \emph{decreasing} values of  ${\rm arg}[ Z_{\gamma_{r_1,r_2}} e^{-i \alpha_0}]$. Thus we have
\begin{equation}
\prod^{\leftarrow}_{\frac{r_1}{r_2} \nearrow} T_{r_1,r_2}^{- \Omega_{r_1,r_2}^+}
\prod^{\leftarrow}_{\frac{r_1}{r_2} \nearrow} T_{r_1,r_2}^{\Omega_{r_1,r_2}^-} = 1
\end{equation}
where  the arrows on the product mean that increasing values of $r_1/r_2$ are written to the left, and
$\Omega_{r_1,r_2}^\pm$ is the BPS index of $r_1 \gamma_1 + r_2 \gamma_2$ in the region $\CU$ with
${\rm Im} Z_1 \bar Z_2 >0$ and $<0$ respectively. Taking into account the relation between the
ordering of $r_1/r_2$ and the ordering of the phases of the central charges illustrated in
figures \ref{fig:Walls} and \ref{fig:CentralCharges} we can also write this in the more traditional
way:
\begin{equation}
\prod^{\rightarrow}_{{\rm arg} Z_{r_1,r_2} \nearrow } T_{r_1,r_2}^{\Omega_{r_1,r_2}^+}=
\prod^{\rightarrow}_{{\rm arg} Z_{r_1,r_2} \nearrow  } T_{r_1,r_2}^{\Omega_{r_1,r_2}^-}  .
\end{equation}
This is the KS wall crossing formula.

We still need to fill in a gap above and
 justify the important claim that we can choose $\gamma_1, \gamma_2$ so that
only $r_1, r_2$ both $\geq 0$ or $\leq 0$ are populated. This ``root basis property''
can   be rigorously proven in certain field theory examples
\cite{Gaiotto:2009hg}. We offer an alternative   justification here
 by requiring that   the spectrum of
BPS masses should not have an accumulation point at zero. (We are therefore using that
the point $t_{ms}$ is not at a singularity of moduli space since that assumption is violated, for example
at a superconformal point.)  Denoting the  central charges
of $\gamma_1,\gamma_2$ at $t_{ms}$ by $\rho_1, \rho_2$ we therefore know that there is an
$\epsilon>0$ so that populated charges $\gamma_{r_1,r_2}$ must have
$\vert r_1 \rho_1 + r_2 \rho_2 \vert > \epsilon$. In the $(r_1,r_2)$-plane this is a strip
of width $2 \epsilon$ centered on  the line with slope $-\rho_2/\rho_1$. (Since $t_{ms}$ is on the marginal stability wall
 $\rho_2/\rho_1$ is real.) If our point $t_{ms}$ is generic then there is in fact a
neighborhood of $t_{ms}$ in the marginal stability wall so that, moving along this wall the spectrum
of BPS particles of charges of the form  $\gamma_{r_1,r_2}$ must remain constant. But the
slope $-\rho_2/\rho_1$ will vary in this neighborhood. This means that there must be an
unpopulated wedge (and its negative)  in the $(r_1,r_2)$-plane. By choosing a suitable
redefinition $\gamma_1 \to  a \gamma_1 + b \gamma_2, \gamma_2 \to c \gamma_1 + d \gamma_2$
we can ensure that the populated states in the complementary wedges are of the form
$\gamma_{r_1,r_2}$ with $r_1, r_2$ both $\geq 0$ or both $\leq 0$.

We end with two remarks

\begin{enumerate}

\item
The root basis property of BPS states is addressed in the mathematical framework of Kontsevich and Soibelman   \cite{KS-1} in a slightly different way. A part of their ``stability conditions,'' used   a quadratric form on the lattice of charges $\CQ: L \to \IR$ and only the charges that satisfied $\CQ(\gamma)\ge 0$ were considered. The quadratic form also has the property that $\CQ|_{{\rm Ker} \, Z}<0$, where $Z$ is the central charge map $Z: L \to \IC$. Thus, restricting the set of charges entering the WCF to $\CQ(\gamma)\ge 0$  means that  we have to discard certain wedges in the space of charges surrounding the directions with $Z(\gamma)=0$.

\item
Finally, we comment on the ``motivic'' or ``refined'' version of the wall-crossing formula \cite{KS-1}
which takes into account spin degrees of freedom \cite{Dimofte:2009bv,Dimofte:2009tm}.
The field theoretic derivation of the motivic KSWCF  given in \cite{Gaiotto:2010be} can also
be carried over directly in the present context: We now let $X_\gamma$ be valued in the quantum
torus. We replace $\F_{C}$ by the generating function of the spin characters,
 and across the walls $W_{\gamma}$ we will find that $\F_{C}$ is conjugated by
  certain combinations of quantum dilogarithms.
However, we stress that the justification for the derivation in   \cite{Gaiotto:2010be}    relied
on the existence of ``\emph{protected spin characters},'' which can only be defined if there is   an $SU(2)_R$ symmetry
in the supersymmetry algebra. In general this symmetry is not present in supergravity, and hence,
from the viewpoint of this Letter, the validity of ``motivic'' generalization of the
wall-crossing formula is a little mysterious. In fact, as is well-known,   the spin character depends on hypermultiplet moduli as well as vectormultiplet moduli. (For examples in the 
weakly coupled heterotic strings with type II duals see \cite{Harvey:1995fq,Dabholkar:2005by}.)

\end{enumerate}

\section{Generalization to noncontractible loops} \label{sec:noncontr}

In our derivation of the KS formula, we considered a contractible loop $\CP$ in moduli space. Nothing prevents us from considering instead a non-contractible loop, in particular a loop circling around a point on the discriminant locus. Such a loop will be closed in moduli space but not in covering space, and the local system of charges undergoes nontrivial monodromy $M_{\CP}: L \to L$ after going around it.\footnote{To avoid cluttering the discussion, in the following we will not bother specifying at each step in which direction we orient loops, monodromies etc.} As a result the generating function will not be exactly preserved, and (\ref{eq:trivmonF1}) must be modified.

As mentioned under (\ref{eq:gen-fun}), the proximity of singularities associated to nontrivial monodromies can lead to some subtleties in the definition of the framed BPS indices $\fro_C(\Gamma_{\rm orb},t_\infty)$. Besides the usual jumps at marginal stability, there are two other kinds of formal index ``jumps'' (or rather relabelings) related to the presence of singularities and monodromies. The first occurs when $t_\infty$ crosses a cut, where the choice of charge lattice basis jumps by convention. This is just a relabeling of indices, equating framed indices involving charges related by the corresponding basis transformation. If desired it can be eliminated by going to the moduli covering space. The second event occurs when $t_\infty$ crosses a ``conjugation wall'' in the language of \cite{BST}, i.e.\ when the core attractor flow gets ``pulled through'' a singular locus in moduli space. In this case new particles (becoming massless at the singularity) appear in orbit while the apparent core charge as seen from infinity jumps, keeping the total charge (and index) unchanged. This is again some kind of relabeling of indices, equating framed indices involving shifted core and orbit charges, but this time the jump cannot be eliminated by going to the covering space.

More formally, when crossing a cut from $t_\infty$ to $t_\infty'$, charges $\Gamma|_{t_\infty'}$ and $M \cdot \Gamma|_{t_\infty}$ get identified. Thus the indices on the respective sides of the cut are related by
\begin{equation}
 \fro_{C}(\Gamma_{\rm orb};t_\infty') = \fro_{M\cdot C}(M \cdot\Gamma_{\rm orb};t_\infty) \, .
\end{equation}
A short computation shows that the generating functions get accordingly identified as
\begin{equation} \label{cutmon}
 \F_{C}(X;t_\infty) = \hat{M} \cdot \F_{M^{-1} C}(X,t_\infty') \, ,
\end{equation}
where we defined for any automorphism $M$ of the charge lattice a map on generating functions by
\begin{equation} \label{KMdef}
 \hat{M} \cdot \sum_\Gamma a_\Gamma X^\Gamma := \sum_\Gamma a_\Gamma X^{M \cdot \Gamma} \, . \end{equation}

When crossing a conjugation wall from $t_\infty$ to $t_\infty'$, by definition, the core attractor flow gets pulled through the discriminant locus, so that if initially the core attractor flow did not cross the cut ending on the discriminant locus, it now does, or vice versa. By physical continuity, the core charge as seen by a local observer at the core must remain $\Gamma_c$. Hence, if the monodromy transformation associated to the cut is $\Gamma \to M \cdot \Gamma$, the apparent core charge as seen by an observer at spatial infinity jumps from $\Gamma_c$ to $M \cdot \Gamma_c$. Since the total charge must remain the same, the charge in the galactic orbit must jump from $\Gamma_{\rm orb}$ to $\Gamma_{\rm orb} + (1-M) \cdot \Gamma_c$ (see \cite{BST} for a detailed discussion of how this happens physically). Note that to remain in the picture in which the orbit charge remains finite when $\Lambda \to \infty$, we should therefore require
\begin{equation} \label{invGamma0}
 M \cdot \Gamma_0 = \Gamma_0 \, , \qquad M \cdot \Gamma_0' = \Gamma_0' \, .
\end{equation}
The framed indices on the respective sides of the conjugation wall are then related by
\begin{equation}
 \fro_{C}(\Gamma_{\rm orb};t_\infty) = \fro_{M \cdot C}(\Gamma_{\rm orb}+(1-M)\cdot \gamma_c;t_\infty') \, .
\end{equation}
The corresponding generating functions are related even more simply by
\begin{equation}
 \F_{C}(X,t_\infty) = \F_{M\cdot C}(X,t_\infty') \, .
\end{equation}

We can now collect these results and state the generalization of (\ref{eq:trivmonF1}) to the case of a noncontractible loop $\CP$ around a point $t_0$ of the discriminant locus, with associated monodromy $M$. As before, we assume that no massless BPS particles exist at $t_*(\Gamma_0)$. Since in general there are massless BPS particles present at the discriminant locus, we assume in particular that we have chosen $\Gamma_0$ to be such that $t_0 \neq t_*(\Gamma_0)$. There are two cases to distinguish:
\begin{enumerate}
 \item {\bf Singularity without conjugation wall}: This is the case for singularities at infinite distance, such as the infinite volume limit of IIA on the quintic. We can assume there is a single cut ending on the singularity, across which the generating function transforms as in (\ref{cutmon}). Going infinitesimally across the cut in one direction or along the full loop $\CP$ in the other direction (along which the generating function undergoes a series of wall crossing operations as before), should give the same result. Thus (\ref{eq:trivmonF1}) generalizes to
\begin{equation}
   \prod_i U_{\gamma_i}(t_i) \cdot \F_{C}   = \hat{M} \cdot \F_{M^{-1} \cdot C} \, .
\end{equation}
\item {\bf Singularity with conjugation wall}: This is the case typically for singularities at finite distance, such as the conifold point of IIA on the quintic. Assuming (\ref{invGamma0}) and taking without loss of generality the cut on top of the conjugation wall for convenience, the transformation of the generating function when crossing the wall is given simply by $\F_{C} \to \F_{M \cdot C} \to \hat{M} \cdot \F_{C}$, and the analog of (\ref{eq:trivmonF1}) becomes
\begin{equation}
   \prod_i U_{\gamma_i}(t_i) \cdot \F_{C}   = \hat{M} \cdot \F_{C} \, .
\end{equation}
By the same arguments as before, we can infer from this the operator equation
\begin{equation} \label{nontrivmoneq}
 \prod_i U_{\gamma_i}(t_i)  = \hat{M} \, ,
\end{equation}
which generalizes (\ref{eq:trivmon}).

As an application of this formula, consider a singularity $t_0$ where a charge $\gamma$ becomes massless, but no other linearly independent charges do.
Because $Z(\gamma)$ acquires all phases around $t_0$, the loop $\CP$ will necessarily cross both $W_\gamma$ and $W_{-\gamma}$. If the loop is chosen such that these are the only walls that are crossed, equation (\ref{nontrivmoneq}) becomes
\begin{eqnarray}
 \hat{M} &=& U_{-\gamma} \cdot U_{\gamma} \nonumber \\
 &=&
 \prod_{k} \left(1 - (-1)^{-k D_{\gamma}} X^{-k \gamma} \right)^{- k \Omega(k \gamma) D_{\gamma}}
 \prod_{k} \left(1 - (-1)^{k D_{\gamma}} X^{k \gamma} \right)^{k \Omega(k \gamma) D_{\gamma}}
 \nonumber \\
 &=& X^{\sum_k k^2 \Omega(k \gamma) \, \gamma D_\gamma} \, .
\end{eqnarray}
Recalling (\ref{KMdef}), we see this is equivalent to
\begin{equation}
 M \cdot \Gamma = \Gamma + \sum_k k^2 \Omega(k \gamma) \, \langle \gamma,\Gamma \rangle \, \gamma \, .
\end{equation}
Thus this generalized KS formula relates monodromy to the the BPS spectrum. In the case of the simple conifold, $\Omega(k \gamma) = \delta_{k,0}$ and the above formula reduces to the well know Picard-Lefshetz monodromy formula $M \cdot \Gamma = \Gamma + \langle \gamma,\Gamma \rangle\, \gamma$. We discuss such relations in much more detail in \cite{BST}.

\end{enumerate}

\begin{section}* {Acknowledgements}

We would like to thank Dionysis Anninos, Miranda Cheng, Clay Cordova, Davide Gaiotto and
Andy Neitzke  for   discussions.   This work is supported by the DOE under grants
DE-FG02-96ER40959 and DE-FG02-91ER40654. GM would like to thank the Aspen Center for Physics for hospitality during the completion of this work.

\end{section}

\appendix

\section{No-mixing conditions}\label{sec:MultipleCore}

A crucial element in our derivation of the KS wall crossing formula and its generalization was the argument for the absence of quantum mixing between galaxies with different core charges, and between galaxies with orbit charges $\gamma \in L_{\rm orb}$ and galaxies with some orbit charges $\gamma \notin L_{\rm orb}$. As promised we will now examine this argument in more detail, and show that mixing is absent in the $\Lambda \to \infty$ limit except if there exist massless charged particles at the attractor point of $\Gamma_0$, with charge in $L_{\rm orb}$.

We first investigate nonperturbative quantum mixing between the perturbative semiclassical states corresponding to a galaxy with all orbiting charges $\gamma \in L_{\rm orb}$, i.e.\ $\langle \gamma,\Gamma_0 \rangle = 0 = \langle \gamma,\Gamma_0' \rangle$, and those corresponding to a galaxy with some orbiting charges $\gamma \notin L_{\rm orb}$. The core charge is $\Gamma_c=\Lambda^2 \Gamma_0 + \Lambda \Gamma_0' + \gamma_c$ for both galaxies. This kind of mixing could in principle be mediated by a tunneling process in which a charge $\gamma$ in orbit splits into a charge $\gamma_1 + \delta$ and a charge $\gamma_2 - \delta$, with $\gamma_1,\gamma_2 \in L_{\rm orb}$, $\gamma_1+\gamma_2 = \gamma$, and $\delta \notin L_{\rm orb}$, followed by tunneling of the two charges to their respective BPS equilibrium positions. If the charges are held fixed in the $\Lambda \to \infty$ limit, then since $\delta \notin L_{\rm orb}$ the symplectic product $\langle \delta,\Gamma_c \rangle$ is at least of order $\Lambda$ and therefore by (\ref{Req}) the distance to which the charges would have to tunnel diverges when $\Lambda \to \infty$. Since tunneling over infinite distances is infinitely suppressed, the amplitude for such a process vanishes in the limit. If on the other hand we allow $\delta$ to grow with $\Lambda$, then in particular for $\delta \equiv \Lambda \Gamma_0 + \Gamma_0'$, it is no longer true that $\langle \delta,\Gamma_c \rangle$ diverges. So for such $\delta$ the tunneling trajectory does not have to be infinitely long. However, such diverging charges carry diverging entropy, and hence, by the arguments we will give below, we get infinite entropic tunneling suppression of the splitting event. An even stronger argument is that BPS configurations containing such charge pairs $(\gamma_1 + \delta, \gamma_2 - \delta)$ actually cannot exist, since in the limit $\Lambda \to \infty$, these two charges are essentially opposite (as they diverge but sum up to a finite fixed charge $\gamma$), so they are essentially each others antiparticles, and it is not possible to have particles and anti-particles at the same time in orbit and still be BPS (since particle annihilation would clearly be energetically favorable). Thus, either way, mixing with galaxies with orbiting charges not in $L_{\rm orb}$ does not occur in the limit $\Lambda \to \infty$.\footnote{\label{whyscaling} The preceding reasoning makes clear why we added the somewhat peculiar term $\Lambda \Gamma_0'$ in (\ref{GamLam}): without it there would be unsuppressed tunneling processes for $\delta \propto \Gamma_0$, and with a term $\Lambda^2 \Gamma_0'$ instead, the there would be unsuppressed tunneling for $\delta \propto \Gamma_0 + \Gamma_0'$. Dropping the $\Lambda \Gamma_0'$ term while adding $\Gamma_0$ to $L_{\rm orb}$ would be an alternative, but then the awkward situation arises that all walls $W_{k \gamma+ m \Gamma_0}$ coincide, spoiling the derivation of the KS formula.}

Now we investigate mixing between different cores. Consider a BPS galaxy with core charge $\Gamma_c = \Lambda^2 \Gamma_0 + \Lambda \Gamma_0' + \gamma_c$ and total orbiting charge $\Gamma_{\rm orb}$, and a galaxy with core charge $\Gamma_c' = \Gamma_c - \delta$ and orbiting charge $\Gamma_{\rm orb}' = \Gamma_{\rm orb} + \delta$. The perturbative semiclassical states corresponding to these classical configurations can mix nonperturbatively through tunneling of a BPS particle of charge $\delta$ between the core black hole and a solar system orbiting the galaxy. We will now argue that such tunneling is infinitely suppressed in the limit $\Lambda \to \infty$, except if $\delta$ lies in $L_{\rm orb}$ and becomes massless at the attractor point $t_*(\Gamma_0)$ of $\Gamma_0$.

The infinite suppression when $\delta \notin L_{\rm orb}$ in the limit $\Lambda \to \infty$ follows by essentially the same arguments as we used above to show the absence of mixing between galaxies with all orbiting charges in $L_{\rm orb}$ and galaxies with some orbiting charges not in $L_{\rm orb}$: charges $\delta \notin L_{\rm orb}$ would either have to tunnel infinitely far (when they are kept finite), or (when $\delta \propto \Lambda \Gamma_0 + \Gamma_0'$) have infinite entropy themselves and give rise to an infinite change in entropy of the core. Either way, tunneling is infinitely suppressed.

When $\delta \in L_{\rm orb}$, the particle can tunnel to finite distance, but tunneling will be infinitely suppressed in the limit due to the fact that the change in entropy of the core is infinite, \emph{except} when the mass $m_\delta = |Z_\delta|$ of $\delta$ vanishes at $t_*(\Gamma_0)$. We first show the steps in the proof of this claim and then explain them. The entropy difference is
\begin{eqnarray}
 \Delta S &=& S_{\rm BH}(\Lambda^2 \Gamma_0 + \Lambda \Gamma_0' + \gamma_c) - S_{\rm BH}(\Lambda^2 \Gamma_0 + \Lambda \Gamma_0' + \gamma_c - \delta)  \\
 &=& \Lambda^4 \left[ \frac{\delta^I}{\Lambda^2} \frac{d}{d\Gamma^I} S(\Gamma)|_{\Gamma=\Gamma_0} + \CO(\frac{1}{\Lambda^4}) \right] \\
 &=& \pi \Lambda^2 \delta^I \frac{d}{d\Gamma^I} |Z(\Gamma,t_*(\Gamma))|^2|_{\Gamma=\Gamma_0} + \CO(1) \\
 &=& 2 \pi \Lambda^2 \, {\rm Re}(\overline{Z}_{\Gamma_0} Z_\delta)|_{t_*(\Gamma_0)} + \CO(1) \\
 &=& \pm 2 \pi \Lambda^2 |Z_{\Gamma_0}| \, m_\delta \, |_{t_*(\Gamma_0)} + \CO(1) \, ,
\end{eqnarray}
which indeed diverges when $\Lambda \to \infty$ except if $m_\delta |_{t_*(\Gamma_0)} = 0$. In going from the first to the second line we used the fact that the Bekenstein-Hawking entropy of a BPS black hole scales quadratically with the charges, and we expanded around $\Lambda=\infty$. In the third line we used the expression of the entropy in terms of the central charge. In going to the next to last line we were allowed to ignore the dependence on $\Gamma$ through $t_*(\Gamma)$ because $|Z(\Gamma,t)|$ has a critical point at $t=t_*(\Gamma)$, i.e.\ $\partial_t |Z(\Gamma,t)||_{t_*(\Gamma)}=0$. In the final step we used the attractor point equation $2 \, {\rm Im}(\overline{Z}_{\Gamma_0} Z_\delta)|_{t_*(\Gamma_0)} = \langle \Gamma_0,\delta \rangle = 0$. Thus, in the absence of massless BPS particles at $t_*(\Gamma_0)$ with charge in $L_{\rm orb}$, there can be no mixing between galaxies with different core charges.

In conclusion, if no massless charged particles exist at $t_*(\Gamma_0)$, our BPS galaxies are closed quantum systems in the limit $\Lambda \to \infty$, and the framed index is well defined. Massless charged particles only appear at loci of complex codimension 1. Thus, for a generic $\Gamma_0$, there will be no massless charged particles at $t_*(\Gamma_0)$, and there will be no mixing.

There might however be special circumstances in which we are interested precisely in the situation where $L_{\rm orb}$ contains charges becoming massless at $t_*(\Gamma_0)$. In this case, mixing may occur, so to be guaranteed a well-defined index we should sum over values of the core charge differing by multiples of the charges becoming massless. It is indeed natural to consider such nongeneric situations in compactifications with codimension 1 loci of enhanced gauge symmetry, as we now explain. Near such loci, there are light vector bosons, say of charge $\gamma$, and typically also light monopoles of charge $\gamma_D$. Their central charges are related by $Z_{\gamma_D} \sim \tau Z_\gamma$, where $\tau$ is the (moduli-dependent) complexified coupling, and $Z_\gamma \to 0$ at the enhanced symmetry locus. When we want to allow both the vector boson and the monopole in a galactic orbit, i.e.\ $\gamma,\gamma_D \in L_{\rm orb}$, the attractor equations for $\Gamma_0$ imply ${\rm Im}(Z_\gamma \overline{Z}_{\Gamma_0}) = 0 = {\rm Im}(Z_{\gamma_D} \overline{Z}_{\Gamma_0})$ at $t_*(\Gamma_0)$. Given the relation $Z_{\gamma_D} \sim \tau Z_\gamma$ and ${\rm Im} \, \tau \neq 0$, this implies $Z_\gamma|_{t_*(\Gamma_0)} = 0$ --- that is, we necessarily have massless particles at the attractor point. In Section 7 of \cite{BST} we discuss an example of this sort, and show explicitly that it is indeed necessary to sum over core charges to get a well-defined index.

To make this more precise, we could try to define a generalized framed index by summing over the entire lattice $V_0$ of multiples of charges in $L_{\rm orb}$ becoming massless at $t_*(\Gamma_0)$:
\begin{equation} \label{eq:gen-framed-deg}
\fro_{C}(\Gamma_{\rm orb};t_\infty) := \sum_{\nu \in V_0} \lim_{\Lambda \to \infty} {\rm Tr}_{\CH_{\Gamma_c + \nu}(\Gamma_{\rm orb}-\nu;t_\infty)} \, (-1)^F \, .
\end{equation}
There is some redundancy among these objects, as $\fro_{\{ \Gamma_0,\Gamma_0',\gamma_c \}}(\Gamma_{\rm orb}) = \fro_{\{\Gamma_0,\Gamma_0',\gamma_c+\nu\}}(\Gamma_{\rm orb}-\nu)$ for any $\nu \in V_0$. Consequently, the associated generating function
\begin{equation}
\F_{[C]}(X;t_\infty):= \sum_{\Gamma_{\rm orb}\in L_{\rm orb}} \fro_{C}(\Gamma_{\rm orb};t_\infty) \, X^{\gamma_c + \Gamma_{\rm orb}} \, ,
\end{equation}
depends only on the equivalence class $[C] := \{\Gamma_0,\Gamma_0',\gamma_c \, {\rm mod} \, V_0 \}$. We could now try to repeat the analysis of the previous sections using these generalized definitions. It is however not immediately obvious that the objects we have defined here are finite or computable in practice, and indeed it is only
 in special cases possible to restrict the sum over cores to a finite subset. We will not attempt to provide a general analysis here, but refer to section 7 of \cite{BST} for the study of specific examples.


\begin{thebibliography}{1}


\bibitem{BST}
  E. ~Andriyash, F. ~Denef, D.~ Jafferis, and G.W.~Moore,
  ``Bound State Transformation Walls,''
  to appear.


\bibitem{Dabholkar:2005by}
  A.~Dabholkar, F.~Denef, G.~W.~Moore and B.~Pioline,
  ``Exact and Asymptotic Degeneracies of Small Black Holes,''
  JHEP {\bf 0508}, 021 (2005)
  [arXiv:hep-th/0502157].



\bibitem{Denef:2000nb}
  F.~Denef,
  ``Supergravity flows and D-brane stability,''
  JHEP {\bf 0008}, 050 (2000)
  [arXiv:hep-th/0005049].


\bibitem{Denef:2002ru}
  F.~Denef,
  ``Quantum quivers and Hall/hole halos,''
  JHEP {\bf 0210}, 023 (2002)
  [arXiv:hep-th/0206072].

\bibitem{Denef:2007vg}
  F.~Denef and G.~W.~Moore,
  ``Split states, entropy enigmas, holes and halos,''
  arXiv:hep-th/0702146.


\bibitem{Dimofte:2009bv}
  T.~Dimofte and S.~Gukov,
  ``Refined, Motivic, and Quantum,''
  arXiv:0904.1420 [hep-th].

\bibitem{Dimofte:2009tm}
  T.~Dimofte, S.~Gukov and Y.~Soibelman,
  ``Quantum Wall Crossing in N=2 Gauge Theories,''
  arXiv:0912.1346 [hep-th].





\bibitem{Gaiotto:2009hg}
  D.~Gaiotto, G.~W.~Moore and A.~Neitzke,
  ``Wall-crossing, Hitchin Systems, and the WKB Approximation,''
  arXiv:0907.3987 [hep-th].


\bibitem{Gaiotto:2010be}
  D.~Gaiotto, G.~W.~Moore and A.~Neitzke,
  ``Framed BPS States,''
  arXiv:1006.0146 [hep-th].

\bibitem{Harvey:1995fq}
  J.~A.~Harvey and G.~W.~Moore,
  ``Algebras, BPS States, and Strings,''
  Nucl.\ Phys.\  B {\bf 463}, 315 (1996)
  [arXiv:hep-th/9510182].


\bibitem{KS-1}
M. Kontsevich and Y. Soibelman, ``Stability structures, motivic
Donaldson-Thomas invariants and cluster transformations,''
arXiv:0811.2435

\bibitem{KS-2}
M. Kontsevich and Y. Soibelman, ``Motivic Donaldson-Thomas
invariants:summary of results,'' arXiv:0910.4315

\bibitem{Maldacena:1998uz}
  J.~M.~Maldacena, J.~Michelson and A.~Strominger,
  ``Anti-de Sitter fragmentation,''
  JHEP {\bf 9902}, 011 (1999)
  [arXiv:hep-th/9812073].


\end{thebibliography}
\end{document}